\documentclass[preprint,showpacs,preprintnumbers,amsmath,amssymb]{revtex4}

\usepackage{graphicx}
\usepackage{dcolumn}
\usepackage{bm}
\usepackage{natbib}

\begin{document}

\preprint{To appear, Physica A}

\title{Computational Models of Adult Neurogenesis}

\author{Guillermo A. Cecchi}
 \affiliation{T.J. Watson IBM Research Center,
 Yorktown Heights, NY, USA }
\author{Marcelo O. Magnasco}
\affiliation{The Rockefeller University, New York City, NY, USA}
\date{\today}

\begin{abstract}
Experimental results in recent years have shown that adult
neurogenesis is a significant phenomenon in the mammalian brain.
Little is known, however, about the functional role played by the
generation and destruction of neurons in the context of and adult
brain. Here we propose two models where new projection neurons are
incorporated. We show that in both models, using incorporation and
removal of neurons as a computational tool, it is possible to
achieve a higher computational efficiency that in purely static,
synapse-learning driven networks. We also discuss the implication
for understanding the role of adult neurogenesis in specific brain
areas.
\end{abstract}

\maketitle

\section{Adult Neurogenesis and Computation}

Adult neurogenesis (AN), that is, the incorporation of new neurons
in adult brains, has been documented in the mammalian olfactory
bulb and the dentate gyrus, as well as in in a variety of
subcortical areas and different areas of the neocortex, but with
less significance \cite{ALTMAN, NOTTE, ARTURO}. A number of
observations point to a functional role for AN, in particular the
increase in proliferation after exposure to novel environments
\cite{NOVELDG, NOVELOB}, and the incorporation of new neurons as
functionally connected elements \cite{LEOPOLDO2}. We have recently
introduced the first computational model of adult neurogenesis in
the olfactory bulb \cite{ANOB}, showing that the death and
incorporation of inhibitory (granule) interneurons is enough to
achieve the orthogonalization of an input ensemble, a task the
olfactory bulb is assumed to be involved in \cite{friedrich}. The
model predicted an initial wave of massive cell death for newly
incorporated neurons, followed by a constant, small and protracted
background of death. Interestingly, these qualitative predictions
were subsequently confirmed by experiments \cite{LEOPOLDO1}. This
model, however, was not able to account for the possible
computational necessity of AN, given that similar results can be
obtained by synaptic learning. We are nevertheless interested in
proving the hypothesis that AN is a required solution to specific
problems found by evolution, as opposed to an evolutionary vagary.
We present in this letter two models that provide evidence for
this hypothesis. We will discuss first a model in which the
replacement of neurons is based on their level of activity,
displaying interesting information-theoretic properties. We will
later introduce a model in which the replacement is based on a
measure of correlation between neighboring neurons; this second
model shows interesting properties related to the convergence to
minimal distortion of neural maps.

\section{Replacement models}

The activity-based replacement model consists of a network of
projection units that learn to represent an input ensemble based
on a winner-take-all approach (WTA). This model is related to the
Growing Cell Structure (GCS) algorithm introduced by Fritzke
\cite{FRITZKE}, but unlike it has a simple physiological
interpretation. The GCS approach requires the incorporation of new
neurons with specific synaptic weights, derived from those of
already incorporated neurons; in our algorithm, this
"teleological" requirement is not present.

The training proceeds as follows: upon presentation of an input
${\mathbf X}_n$, a winner unit is selected such that, ${\mathbf
W}_k = \min_i |{\mathbf W}_i - {\mathbf X}_n|$. The winner and the
runner-ups are updated following a standard Hebbian rule: $ \Delta
{\mathbf W}_k \propto {\mathbf X}_n - {\mathbf W}_k$, and $\Delta
{\mathbf W}_i \propto ({\mathbf W}_i - {\mathbf W}_k) f(|{\mathbf
W}_k-{\mathbf W}_i|)$, where $f(\dot)$ is a non-linear decreasing
function of its argument; here we will choose and exponential. We
introduce here the novel component of the algorithm. First, a
winning rate is computed for each unit, $w_i^{n}=1$ if $i=k$, $0$
otherwise, $\Rightarrow \omega_i=\langle w_i\rangle_n$, intended
to capture the unit's recent activity. Based on this value, the
units are replaced. This is done probabilistically,
$p($death$)_i\propto 1-\omega_i$. At the same time, new neurons
arrive constantly, with a probability $p(N+1) = \lambda$. The
newly arrived units are not immediately connected: they are
allowed to participate of the competition process, but do not
become part of the representation ensemble until a trial period
$\tau$ has passed. In this way, the number of output units is
always bounded, and new neurons are allowed to find a good
configuration.

The results of the simulation are shown in Fig 1-A,B. Different
input ensembles consisting of $100$ elements are drawn from a
uniform distribution on the unit $100D$ hyper-sphere. The network
is initialized with an arbitrary number of output units
(50[$\circ$] and 150[$\circ$] in this case), and is trained with a
random sequence of input exemplars. As small amount of noise is
added to the input exemplars, but this is not a determining
factor. Panel A shows the evolution of the mutual information
between the input and the output ensembles, relative to the
maximal mutual entropy (the entropy of the input in this case).
The mutual information here is measured as in a discrete channel,
using only the output units, as $H({\mathbf X},{\mathbf
W})=H({\mathbf W})-H({\mathbf W}|{\mathbf X})$. We observe that
indeed the network evolves towards maximal mutual information, for
both low and high initial number of units. This is expected of a
network implementing the Hebbian component of the algorithm. Panel
B is however more surprising: it shows that the network evolves
towards a number of units equal to the number of input clusters,
independently of the initial number of units. This is indeed a
robust feature, independent also of the total number of input
clusters.


\input{psfig.sty}
\begin{figure}

\psfig{figure=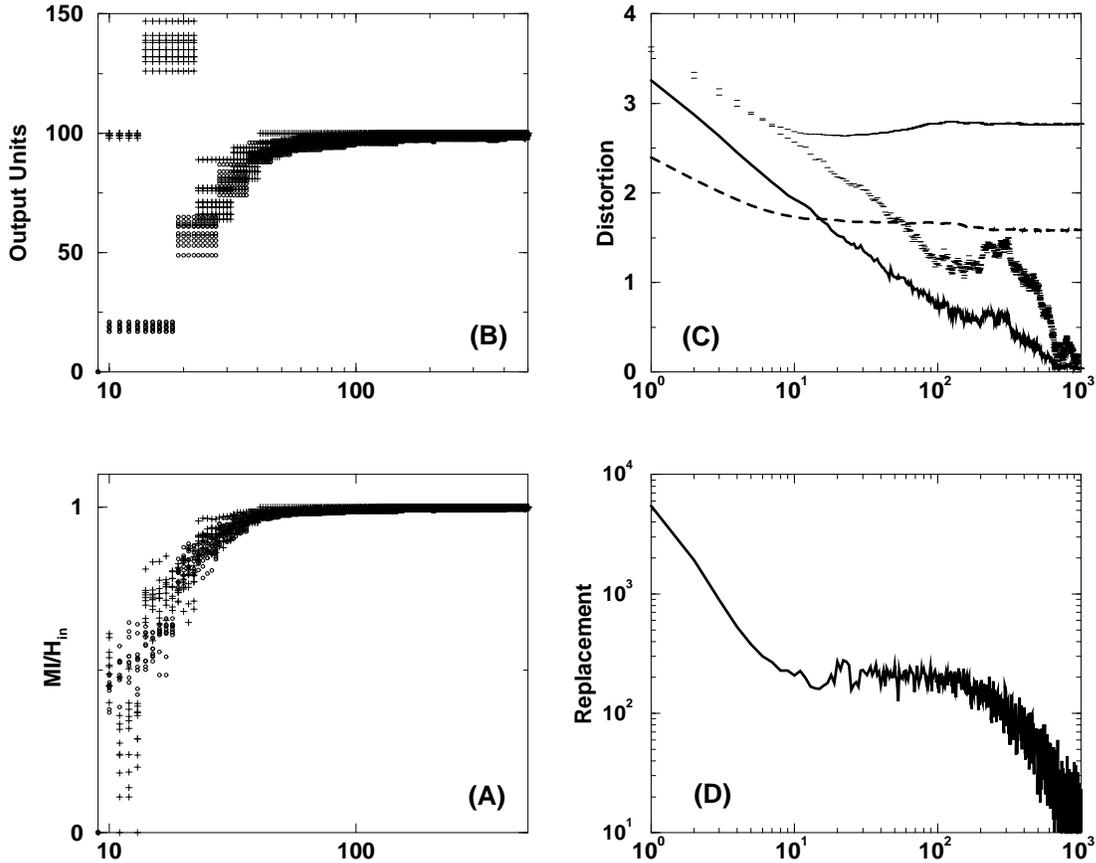,width=6.0in,angle=270}

\caption{A: Mutual Information of activity-based model. B: Total
number of output units in the same model; x-axes are iterations
x$10^3$. C: Distortion averaged over initial configurations for
correlation-based model (solid line), and simple elastic model
(dotted line). Dashes are one standard deviation, and the x-axis
is iterations. D: Average replacement rate for the
correlation-based model}

\label{fig:fig1}
\end{figure}


Given that mutual information maximization can be equally achieved
by a non-modifying Hebbian network, what is the advantage of
implementing a network like this one? A key point here is the
ability of the network to track the number of input clusters in
the number of output units. A simple analysis of a the discrete
channel defined by the network shows that, for a given number of
inputs $M$ and network size $N$, the maximal mutual information
normalized by the input entropy scales like $\log N$ for $N<M$,
and is $1$ when $N \geq M$. At the same time, the redundancy of
the output, defined as $H({\mathbf W}|{\mathbf X})$ is $0$ when
$N<M$, to then scale as $\log N - \log M$. This shows that a
WTA-like network with the same number of units as number of inputs
in the input ensemble (the {\sl cardinality}) can at the same time
maximize the mutual information and minimize the output
redundancy. Both features are relevant in the context of brain
processing \cite{RL}; in particular, minimizing the total number
of outputs without compromising the mutual information can reduce
the wiring and crosstalk in the target network, if at the expense
of maintaining a costly mechanism like AN, and in the context of
an unknown and possibly changing cardinality of the input
ensemble.

The model presented above shows one possible advantage of
combining classical synaptic learning with AN. One limitation of
this model is the the task required from the network is fairly
simple, and lacking therefore generality. In what follows we will
present a different model, in which the replacement of units is
driven by correlated activity, and implementing a less simple form
of computation, i.e. topology-preserving mapping. This
correlation-based replacement model is based on the elastic net
introduced in \cite{DM}. The goal of the network is to represent
the input space in a topographically order fashion, such that
nearby units in neural space have similar receptive
characteristics. The network differs from the previous one in that
no winner needs to be computed, and neighboring units influence
each other. Formally, the update rule is: $\Delta {\mathbf W}_i =
\alpha A_i ({\mathbf X} - {\mathbf W}_i ) + \beta \kappa ({\mathbf
W}_j-{\mathbf W}_i)h(d_{ij})$, where $A_i= e^{-({\mathbf
X}-{\mathbf W}_i)^2/2\kappa^2} / \mathcal{N}$, and $
\mathcal{N}=\sum_j e^{-({\mathbf X}-{\mathbf W}_j)^2/2\kappa^2}$,
can be interpreted as a normalized activity, $d_{ij}$ is the
distance between units $i$ and $j$ in neural space, and $h(\dot)$
a monotonically decreasing function of its argument. In many
cases, as in here, a simpler version is used where only the
immediate first neighbors of each unit are considered. In this
case, the second term of the update equation reads $\beta
({\mathbf W}_{i+1}+{\mathbf W}_{i-1}-2{\mathbf W}_i)$. It is
interesting to notice that the update equation minimizes an energy
function than can be computed by simple integration, $E = - \alpha
\kappa \sum_n \log \sum_i e^{-({\mathbf X_n}-{\mathbf
W}_i)^2/2\kappa^2} - \beta \sum_i ({\mathbf W}_{i+1}-{\mathbf
W}_i)^2$. The elastic net has been used in a variety of
optimization problems, like the travelling salesman problem, and
also in simulations of cortical maps. It is well known, however,
that as many optimization algorithms, the elastic net can get
trapped in local minima depending on the complexity of the input
space. To illustrate this phenomenon, we implemented and elastic
net of $100$ units tasked with learning an input space consisting
of $100$ exemplars distributed on a two dimensional space as $y_i
= 50\sin(x_i\pi /35 + \pi/2 ) + 50$. Without scheduling of
$\kappa$ \cite{DM}, the elastic net can find an optimal solution
only in less the $1\%$ of initial conditions (Fig. 2). In
contrast, the same network, with the addition of the
correlation-based replacement, is able to find the optimal
solution under any initial condition. The algorithm is as follows:
each unit computes a correlation-based or ``stress'' variable,
whose evolution is defined as ${\dot s_i}(t) = \gamma [ \sum_j
({\mathbf W}_i-{\mathbf W}_j)^2 h(d_{ij}) ]^{1/2} - \mu s_i(t)$,
where $\gamma$ and $\mu$ are arbitrary parameters, and the initial
condition for new neurons is $s_i(t)|_{new}=0$. When the
``stress'' reaches a threshold, the unit is ``replaced'' by a new
one with an arbitrary distribution of synaptic weights, $s_i \ge
S_T \Rightarrow$ {\sl reset}$ [{\mathbf W}_i]$. The results of the
simulation are presented in Panel C, where the ensemble average
over initial conditions shows a linearly decreasing behavior in
semilog scale. Although the individual evolution for different
initial configurations can vary dramatically, for our toy example
they all converge to minimal distortion in finite time, in sharp
contrast with the pure elastic net algorithm. Panel D shows the
evolution of the replacement rate, defined as the number of
replaced units per cycle. It displays a fast (power-law) initial
decay, followed by a long exponential tail; interestingly, the
same qualitative features are observed both experimentally and
computationally in the olfactory bulb, as mentioned above.

In summary, we have shown two novel mechanisms that suggest that
AN may be a necessary computational tool used by brain structures
whenever (a) an input ensemble of discrete elements and
non-stationary cardinality needs to be processed (the
activity-based model), or (b) a topographic map of a complex input
space needs to be formed with minimal distortion (the
correlation-based model). The first feature maybe relevant for the
neural processing required by the High Vocal Center, a brain area
of songbrids were replacement correlates with song modification
\cite{ARTURO}. The second feature maybe important for the
generation of an adaptive neural map in the Dentate Gyrus (DG) of
the Hippocampus. We can only speculate at this point, but the
local circuitry of the DG is complex enough to support a
physiological implementation of the correlation-based algorithm
\cite{GORDON}, and the correlation of replacement with exposure to
novel environments \cite{NOVELDG} is compatible with our model.

\bibliographystyle{plain}

\end{document}